\documentclass[aps,pra,floatfix,
twocolumn,showpacs,amsmath,amssymb,superscriptaddress]{revtex4}
\usepackage{times}
\usepackage{latexsym}
\usepackage{graphicx}
\usepackage{verbatim,times,bbm}
\usepackage{color}
\usepackage{appendix}

\newtheorem{lemma}{Lemma}
\newtheorem{Theorem}{Theorem}

\newtheorem{conjecture}{Conjecture}
\newcommand{\proofend}{\hfill\fbox\\\medskip }
\newcommand{\proof}[1]{{\noindent\bf Proof } #1 $\proofend$}

\usepackage{color}

\begin{document}

\title{Minimum output entropy of a non-Gaussian quantum channel}
\author{Laleh Memarzadeh}
\affiliation{Department of Physics, Sharif
University of
Technology, Teheran, Iran}

\author{Stefano Mancini}
\affiliation{School of Science and Technology, University of
Camerino, I-62032 Camerino, Italy}
\affiliation{INFN-Sezione di Perugia, I-06123 Perugia, Italy}

\begin{abstract}
We introduce a model of non-Gaussian quantum channel that stems
from the combination of two physically relevant processes
occurring in open quantum systems, namely amplitude damping and
dephasing.
For it we find input states approaching zero output entropy,
while respecting the input energy constraint.
These states fully exploit the infinite dimensionality of the
Hilbert space. Upon truncation of the latter,
the minimum output entropy remains finite and optimal input
states for such a case are conjectured thanks to
numerical evidences.
\end{abstract}


\date{\today}

\maketitle

\section{Introduction}

Recently the subject of quantum channels has catalysed the
attention for its usefulness in foundational issues as well as
in technological applications (for a recent review, see
\cite{CGLM14}).
Formally a quantum channel is a completely positive and trace
preserving map acting on the set of states (density operators)
living in a Hilbert space. Since any physical process involves a
state change, it can be regarded as a quantum channel mapping
the initial (input) state to the final (output) state. As such
it can be characterized in terms of its information transmission
capability. This implies the use of entropic functionals among
which the minimum output entropy plays a dominant role.
In fact it is related to the minimum amount of noise inherent to
the channel, since it quantifies the minimum uncertainty
occurring at the output of a channel when inputting pure states.
More precisely, the output entropy measures the entanglement of
the input pure state with the environment. Being this latter not
accessible, such entanglement induces loss of quantum coherence
and thus injection of noise at the channel output. Clearly, low
values of entanglement, i.e., of output entropy, correspond to
low communication noise.
As a consequence, the study of output entropy yields useful
insights about channel capacities. In particular, an upper bound
on the classical capacity can be derived from a lower bound on
the output entropy of multiple channel uses \cite{King}.

When studying quantum channels a dichotomy between discrete and
continuous channels usually appears. The formers act on states
living in finite dimensional Hilbert space. In contrast the
latter act on states living in infinite dimensional Hilbert
space. This is reflected in the possibility of using discrete or
continuous variables where to encode classical information.
Among continuous quantum channels attention has been almost
exclusively devoted to Gaussian quantum channels,
that is channels mapping Gaussian input states into Gaussian
output ones \cite{Oleg}. The reason is that they are easily
implementable at experiment level and moreover they also handy
at theoretical level.
For these channels the minimum output entropy was largely
investigated \cite{Vittorio1} and then showed that actually
their classical capacity is achieved through states minimizing
the output entropy \cite{Vittorio2}.

Here, we go beyond the restriction of Gaussianity of continuous
quantum channels and propose a model of non-Gaussian quantum
channel that stems from the combination of two physically
relevant processes that occur in open quantum systems, namely
amplitude damping and dephasing. We then analytically find input
states approaching zero output entropy, while respecting the
input energy constraint. They consist in the superposition of
two number states the farthest away one from the other. In
truncated Hilbert space, we find that beside superposition of
two number states, the so-called binomial states \cite{binomial}
can be optimal depending on the value of channels parameters. We
support this latter results by numerical investigations.

The paper is organized as follows. In Section \ref{sec:mod} we
introduce the model and then we show the existence of
optimal input states achieving zero output entropy in Section \ref{sec:optinput}.
Subsequently, in Section \ref{sec:spacetruncation}, we restrict
our attention to truncated Hilbert space and we conjecture about
the optimality of binomial states, beside superposition of two
number states, and we give numerical evidences of this idea.
Section \ref{sec:conclu} is for concluding remarks.

\vspace{-0.8cm}

\section{The model}\label{sec:mod}

Let us start considering the Hilbert space $L^2(\mathbb{R})$
associated to a single bosonic mode
with ladder operator $a,a^\dag$.

In the framework of dynamical maps,
a typical example of Gaussian process is provided by the
amplitude damping effect
described by the master equation \cite{openq}
\begin{equation*}
\frac{d}{dt}\varrho=2 a \varrho a^{\dag}-a^\dag a\varrho-\varrho
a^\dag a=:\mathcal{L}_{AD}(\varrho),
\end{equation*}
for the density operator $\varrho$.
In contrast, a typical example of non-Gaussian process is
provided by the purely dephasing effect
described by the master equation \cite{openq}
\begin{equation*}
\frac{d}{dt}\varrho=2a^\dag a\varrho a^\dag a -(a^\dag a)^2
\varrho-\varrho (a^\dag a)^2=:\mathcal{L}_{PD}(\varrho).
\end{equation*}
In order to interpolate between these two regimes we are going
to consider the following dynamics
\begin{equation}\label{master}
\frac{d}{dt}\varrho=(1-\epsilon)\mathcal{L}_{AD}(\varrho)+\epsilon\mathcal{L}_{PD}(\varrho),
\end{equation}
with $\epsilon\in[0,1]$. It is easy to see that
\begin{equation*}
\mathcal{L}_{AD}(\mathcal{L}_{PD}(\varrho))=\mathcal{L}_{PD}(\mathcal{L}_{AD}(\varrho)).
\end{equation*}
Therefore we can write the formal solution of \eqref{master} as
\begin{equation}\label{prodmaps}
\varrho(t)=e^{(1-\epsilon)t\mathcal{L}_{AD}}e^{\epsilon
t\mathcal{L}_{PD}}\varrho(0).
\end{equation}
Actually this map can be regarded as a
quantum channel $\Phi_{\epsilon,t}$ (depending on the parameters
$\epsilon$ and $t$) mapping
\begin{equation}\label{channel}
\varrho(0)\mapsto\varrho(t)=\Phi_{\epsilon,t}(\varrho(0))=\sum_{j,k=0}^\infty
E_{jk} \varrho(0) E_{jk}^\dag,
\end{equation}
where $E_{jk}$ are the Karus operators \cite{CGLM14}.
In view of \eqref{prodmaps}
\begin{equation*}
E_{jk}=A_jP_k, 
\end{equation*}
where $A_j$ are the amplitude damping Kraus operators
\cite{Liu04}
\begin{equation}\label{krausA}
A_j=\sum_{l=j}^\infty \sqrt{
l \choose j } \left[ 1-f(\epsilon,t) \right]^{(l-j)/2}
\left[f(\epsilon,t)\right]^{j/2} |l-j\rangle\langle l |,
\end{equation}
with $f(\epsilon,t):=1-e^{-2(1-\epsilon)t}$,
and $P_k$ are the phase damping Kraus operators \cite{Liu04}
\begin{equation}\label{krausP}
P_k=\sum_{l=0}^\infty \sqrt{\frac{(2 l^2 \epsilon t)^k}{k!}}
e^{-l^2 \epsilon t} |l\rangle\langle l|.
\end{equation}
In Eqs.\eqref{krausA} and \eqref{krausP} it is used the Fock
basis $\{ |l\rangle\}_{l\in \mathbb{N}_0}$ representation.
Expanding $\varrho(0)$ in the same basis as 
$\varrho(0)=\sum_{m,n=0}^{\infty}C_{m,n}(0)|m\rangle\langle n|$ and
considering the channel in \eqref{channel}, we obtain
\begin{equation}\label{outexp}
\varrho(t)=\sum_{m,n=0}^{\infty}C_{m,n}(t)|m\rangle\langle n|,
\end{equation}
with
\begin{equation}\label{coeff}
\begin{aligned}
C_{m,n}(t)&=e^{-Y_{m,n}(\epsilon)t}\\
&\times\sum_{l=0}^{\infty}C_{_{m+l,n+l}}(0)
\left[{m+l \choose l}{n+l \choose l}\right]^{\frac{1}{2}} f^l,
\\
\end{aligned}
\end{equation}
in which $Y_{m,n}(\epsilon):=(1-\epsilon)(m+n)+\epsilon (m-n)^2$.
Equation \eqref{coeff} is also the solution of the following
recursive relation
\begin{equation}\label{recursive}
\begin{aligned}
\dot{C}_{m,n}(t)&=2(1-\epsilon)\sqrt{(m+1)(n+1)}C_{m+1,n+1}(t)\\&-Y_{m,n}(\epsilon)C_{m,n}(t),
\end{aligned}
\end{equation}
which is obtainable from the master equation \eqref{master}.

When dealing with quantum channels acting on the set of states
living in an infinite dimensional Hilbert space,
it is customary to 
employ the constraint of fixed average
input energy, that is
\begin{equation}\label{inenergy}
{\rm Tr}\left(\varrho(0) a^\dag a\right)=N.
\end{equation}

\vspace{-0.8cm}
\section{Minimizing output entropy}\label{sec:optinput}

The output entropy of the quantum channel $\Phi$ in
Eq.\eqref{channel} is the von Neumann entropy of the output
state, namely
\begin{equation}
S\left(\Phi_{\epsilon,t}\left(\rho\right)\right) 
:= -{\rm Tr}\left[ \Phi_{\epsilon,t}\left(\rho\right)
\log_2\left(\Phi_{\epsilon,t}\left(\rho\right)\right)\right].
\end{equation}
In order to quantify the noise inherent to the quantum channel
$\Phi_{\epsilon,t}$
we look for its minimal output entropy and call the state with
minimum output entropy the optimal input state.

The following Theorem states the existence of states with zero
output entropy.
\begin{Theorem}\label{Theorem}
Input states 
\begin{equation}\label{kappa}
|\kappa_{\alpha}\rangle=\sqrt{1-\frac{N}{K}}|0\rangle+\sqrt{\frac{N}{K}}e^{i\alpha
K}|K\rangle, \quad K\in\mathbb{N},
\end{equation}
with $\alpha\in\mathbb{R}$, respect the input energy constraint \eqref{inenergy} and satisfy
\begin{equation}
\lim_{K\to\infty}S\left(\Phi_{\epsilon,t}\left(|\kappa_{\alpha}\rangle\langle\kappa_{\alpha}|\right)\right)=0,
\end{equation}
for all values of $\epsilon$ and $t$.
\end{Theorem}

\proof{
First we note that all the states $|\kappa_{\alpha}\rangle$ have the
same output entropy due to the covariance property
of the channel under unitary transformations 
\begin{equation*}\label{calU}
U\in\mathcal{U}:=\bigg\{\sum_n e^{i\alpha n} |n\rangle\langle n|
\,\Big|\, \alpha\in\mathbb{R}, n\in\mathbb{N}_0\bigg\}.
\end{equation*}
Therefore, we prove the theorem for $|\kappa_{0}\rangle$.
Using Eq.\eqref{channel}, the corresponding output reads
\begin{equation*}
\begin{aligned}
&\Phi_{\epsilon,t}\left(|\kappa_0\rangle\langle\kappa_0|\right)=\left(1-\frac{N}{K}(1-f^K)\right)|0\rangle\langle
0|\\
&+\sqrt{\frac{N}{K}\left(1-\frac{N}{K}\right)}(1-f)^{K}e^{-\epsilon
K^2t}
\left(|0\rangle\langle K|+|K\rangle\langle 0|\right)\\
&+\frac{N}{K}\sum_{m=1}^K {K \choose m
}(1-f)^mf^{K-m}|m\rangle\langle m|.
\end{aligned}
\end{equation*}
The matrix form of this output state is block-diagonal, so the
eigenvalues can be easily found as
\begin{equation*}
\begin{aligned}
\lambda_{0,K}&=\frac{1}{2}
\left(A+B\pm\sqrt{(A-B)^2+4C^2}\right),\nonumber\\
\lambda_m&=\frac{N}{K} {K \choose m }(1-f)^mf^{K-m}, \qquad
m=1,\ldots, K-1,
\end{aligned}
\end{equation*}
with 
\begin{equation*}
\begin{aligned}
A&:=1-\frac{N}{K}\left(1-f^K\right),\nonumber\\
B&:=\frac{N}{K}\left(1-f\right)^K,\nonumber\\
C&:=\sqrt{\frac{N}{K}\left(1-\frac{N}{K}\right)}(1-f)^Ke^{-\epsilon
K^2 t}.
\end{aligned}
\end{equation*}
It is easy to see that all the eigenvalues approach zero for
$K\rightarrow \infty$, except $\lambda_0$ that approaches one.
Therefore the input state \eqref{kappa}, while satisfying the
input energy constraint, leads to a zero output entropy. More
precisely, its output entropy results:
\begin{eqnarray}\label{Soutkappa}
&&S\left(\Phi_{\epsilon,t}\left(|\kappa_0\rangle\langle\kappa_0|\right)\right)=-\lambda_{0}\log_2\lambda_{0}
-\lambda_{K}\log_2\lambda_{K}\nonumber\\
&&-\frac{N}{K} \left[1-f^K-(1-f)^K\right]
\log_2\left(\frac{N}{K}\right)\nonumber\\
&&+\frac{N}{K}\left[f^K\log_2(f^K)+(1-f)^K\log_2\left((1-f)^K\right)\right]\nonumber\\
&&+\frac{N}{K}\frac{1}{2}\log_2\left(2\pi e Kf(1-f)\right)+{\cal
O}\left(\frac{1}{K}\right).
\end{eqnarray}
Now fixing $\mathcal{E}>0$ we should find
$\mathcal{K}\in\mathbb{N}$ such that
$S\left(\Phi_{\epsilon,t}\left(|\kappa_0\rangle\langle\kappa_0|\right)\right)<\mathcal{E}$
for $K>\mathcal{K}$.
To this end we first find an upper bound for \eqref{Soutkappa}.
Since projective measurements increase entropy \cite{MOhya}, we
have the inequality
\begin{equation*}
S\left(\Phi_{\epsilon,t}\left(|\kappa_0\rangle\langle\kappa_0|\right)\right)
\le H\left(p_{\epsilon,t}(n)\right)
\end{equation*}
where the r.h.s. is the Shannon entropy of the probability mass
function $p_{\epsilon,t}(n):=\langle
n|
\Phi_{\epsilon,t}\left(|\kappa_0\rangle\langle\kappa_0|\right)
|n\rangle$.
Explicitly the latter reads
\begin{equation*}
p_{\epsilon,t}(n)=\left\{
\begin{array}{ccc}
1-\frac{N}{K}(1-f^K)& &n=0\cr\cr
 \frac{N}{K}{K \choose n }(1-f)^mf^{K-n}& &n=1,\ldots, K
\end{array}
\right..
\end{equation*}
As a consequence
\begin{equation*}
\begin{aligned}
H\left(p_{\epsilon,t}(n)\right)&=-\left[1-\frac{N}{K}\left(1-f^K\right)\right]\log_2\left[1-\frac{N}{K}\left(1-f^K\right)\right]
\nonumber\\
&-\frac{N}{K}\log_2\left(\frac{N}{K}\right)+\frac{N}{K}f^K\log_2\left(\frac{N}{K}f^K\right)\nonumber\\
&+\frac{1}{2}\frac{N}{K}\log_2\left(2\pi e K f(1-f)\right)+{\cal
O}\left(\frac{1}{K}\right).
\end{aligned}
\end{equation*}
Using the inequality $-x\log x<\sqrt{x(1-x)}$ we then get
\begin{equation*}
\begin{aligned}
H\left(p_{\epsilon,t}(n)\right)&\leq \sqrt{
\left(1-\frac{N}{K}\left(1-f^K\right)\right)\frac{N}{K}\left(1-f^K\right)}
\nonumber\\
&+\sqrt{\frac{N}{K}\left(1-\frac{N}{K}\right)}
+\frac{1}{2}\frac{N}{K}\sqrt{2\pi e
Kf\left(1-f\right)}\nonumber\\
&\leq
2\sqrt{\frac{N}{K}}+\sqrt{\frac{N}{K}\frac{\pi}{2}eNf(1-f)}.
\end{aligned}
\end{equation*}
By imposing that the above r.h.s. becomes smaller than $\mathcal
E$, it follows
\begin{equation*}
\mathcal{K}=\left\lceil \frac{N}{{\cal
E}^2}\left(2+\sqrt{\frac{\pi}{2}eNf(1-f)}\right)^2 \right\rceil.
\end{equation*}}

\vspace{-0.8cm}

\section{Space Truncation}\label{sec:spacetruncation}

In the previous Section we showed that the input states
\eqref{kappa} give zero output entropy for
$K\rightarrow\infty$. However, if we truncate the Hilbert space
to a finite value of $K$, it is not guaranteed that these
states are still optimal. Finding the optimal input
state under that condition is the aim of this Section.
To start with, we introduce a class of states knows as binomial
states \cite{binomial}
\begin{equation}\label{Binomial}
|B\rangle_{M,\mu}:=\sum_{n=0}^M\beta_{n}|n\rangle,\;\;\;
\beta_{n}:=\left[{M \choose
n}\mu^n(1-\mu)^{M-n}\right]^{\frac{1}{2}},
\end{equation}
with parameters $M\in\mathbb{N}$ and $\mu\in[0,1]$. The binomial
state \eqref{Binomial} reduces to the number state $|0\rangle$
for $\mu=0$ and to the number state $|M\rangle$ for $\mu=1$.
In contrast, in the limit $\mu\rightarrow 0$,
$M\rightarrow \infty$, and $\mu M=\alpha\in\mathbb{R}$ the
binomial state approaches the coherent state $|\alpha\rangle$.

The energy constraint \eqref{inenergy} yields the relation
\begin{equation*}\label{binave}
{\rm Tr}\left( |B\rangle_{M,\mu}\langle B| a^\dag a\right)=M\mu
= N.
\end{equation*}
Furthermore, inserting the coefficients $\beta_n$ of \eqref{Binomial} into \eqref{coeff} we get
the explicit expression
of the output density operator representation in the Fock basis
\begin{equation*}\label{BinoOut}
\begin{aligned}
&\Phi_{\epsilon,t}(|B\rangle_{M,\mu}\langle
B|)=\sum_{m,n=0}^Me^{-Y_{m,n}(\epsilon)t}
\left(\frac{\mu}{1-\mu}\right)^{\frac{m+n}{2}}\\
&\times \sum_{l=0}^{M-\max\{m,n\}}
\left[ {M \choose m+l} {m+l \choose m}{M \choose n+l} {n+l
\choose n}\right]^{\frac{1}{2}}\\
& \times \left(\mu f \right)^l 
\left(1-\mu\right)^{M-l} |m\rangle\langle n|.
\end{aligned}
\end{equation*}
Here we numerically evaluate the output entropy for binomial
input states with average energy $N$. Once $N$ is fixed we still
have the freedom to vary $\mu$ and $M$ in a way that $\mu M=N$.
Since $\mu\leq 1$, for fixed $N$, we
increase $M$ from $\lceil N \rceil$ to $K$,
in order to find the minimum value of
$S\left({\Phi_{\epsilon,t}\left(|B\rangle_{M,\mu}\langle
B|\right)}\right)$. From here on, when we refer to the binomial state $|B\rangle$,
we mean the one which has minimum output entropy among other
possible binomial states with average energy $N$.

 \begin{figure}[t]
\centering
   \includegraphics[scale=0.4]{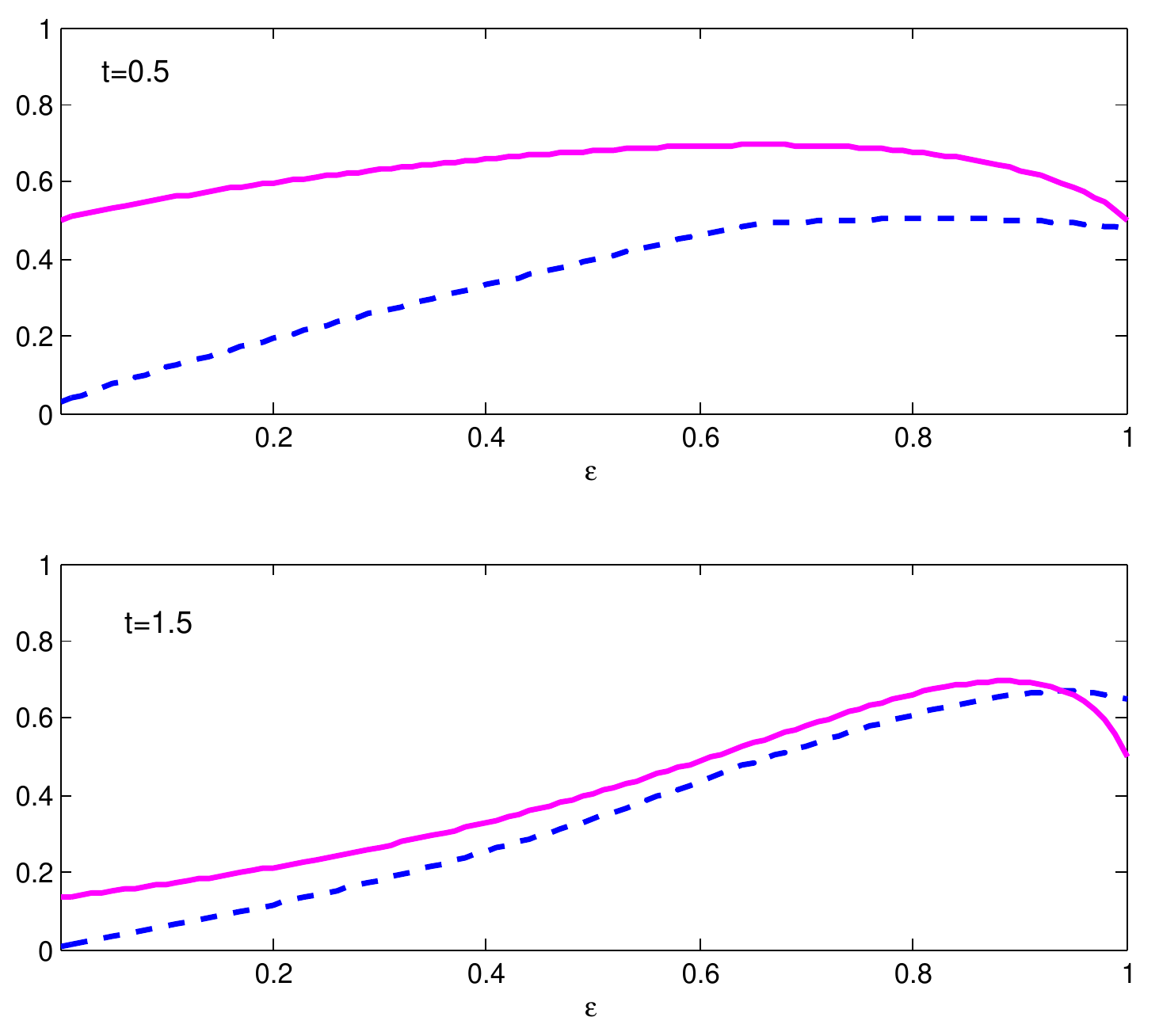}
 \centering
\caption{(Color Online) Output entropy for input state $|B\rangle$ (Blue dashed line) and
for $|\kappa_0\rangle$ (solid magenta line)
 input states with $N=0.6$ ant  $t=0.5$ (top), $t=1.5$ (bottom).}
\label{Bino_Magic}
\end{figure}

\begin{figure}[t]
\centering
   \includegraphics[width=0.4\textwidth]{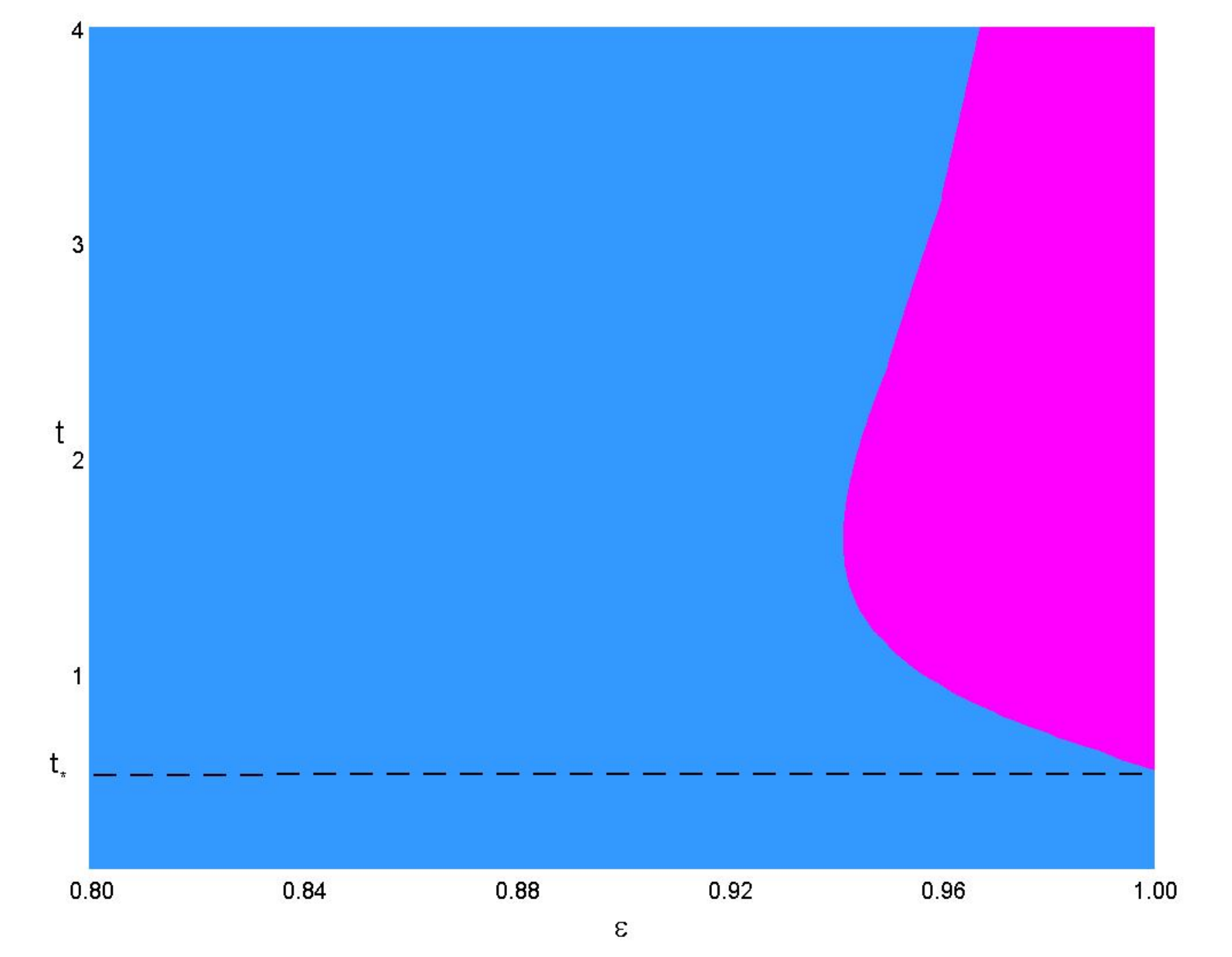}
 \centering
 \caption{ Curve in the $\epsilon,t$ plane where $S(\Phi_{\epsilon,t}(|B\rangle\langle B|))=
 S(\Phi_{\epsilon,t}(|\kappa_0\rangle\langle \kappa_0|))$  for $N=0.6$. On the left (resp. on the right) of the
 curve it is $S(\Phi_{\epsilon,t}(|B\rangle\langle B|))<
 S(\Phi_{\epsilon,t}(|\kappa_0\rangle\langle \kappa_0|))$
 (resp. $S(\Phi_{\epsilon,t}(|B\rangle\langle B|))>
 S(\Phi_{\epsilon,t}(|\kappa_0\rangle\langle \kappa_0|))$).
 The horizontal dashed line represents the value of $t_*$.
 }
\label{tstar}
\end{figure}

Figure \ref{Bino_Magic} shows the output entropy of the state
$|B\rangle$ in \eqref{Binomial} (Blue dashed line)
and of the state $|\kappa_0\rangle$ in \eqref{kappa}
(Magenta solid line) versus $\epsilon$ for $N=0.6$ at $t=0.5$
(top) and $t=1.5$ (bottom). Here $4$-dimensional Hilbert space is
considered. As can be argued from these figures, the output entropy
of $|B\rangle$ remains smaller than the output entropy of
$|\kappa_0\rangle$ (for any value of $\epsilon$)
until $t$ reaches a threshold $t_*$. Then, for $t>t_*$ the state with less
output entropy can be either $|B\rangle$ or $|\kappa\rangle$
depending on the value of $\epsilon$ (see also Fig.\ref{tstar}).

To have an estimation of $t_*$, we first point out that our
numerical analysis shows that the output entropy of
$|B\rangle$ and $|\kappa_0\rangle$ cross each other at large
values of $\epsilon$ where the optimal value of $M$ is 1. In such a case
the output state of $|B\rangle$ lives in a two
dimensional subspace and its output entropy turns out to be
\begin{equation*}
\begin{aligned}
S_B&=-\sum_{j=1}^2\mu_j\log(\mu_j),\\
\mu_{1,2}&:=\frac{1\pm\sqrt{(1-2N(1-f))^2+4N(1-N)e^{-2t}}}{2}.
\end{aligned}
\end{equation*}
Then solving the equation
$S(\Phi_{\epsilon,t}\left(|\kappa_0\rangle\langle\kappa_0|)\right)=S_B, $
we can find the value of $t_*$.  

To do the similar calculation for any given $N$, we have
numerically found that the optimal value of $M$ is
$\lceil N\rceil$. Therefore the output entropy of
$|B\rangle_{M,\mu}$, with $M=\lceil N\rceil$ and $\mu=N/M$,
should be found and equated to
$S(\Phi_{\epsilon,t}\left(|\kappa_0\rangle\langle\kappa_0|)\right)$
in order to get $t_*$.

After having compared the behaviour of the output entropy for
inputs of the kind \eqref{kappa} and \eqref{Binomial}, we
formulate the following conjecture.

\begin{conjecture}\label{conj:bin+magic}
In a truncated Hilbert space of dimension $K+1$, the
minimal output entropy of the quantum channel \eqref{channel} is
achieved either by binomial states of Eq.\eqref{Binomial} or by
states $|\kappa_{\alpha}\rangle$ of Eq. \eqref{kappa},
depending on the values of $\epsilon$ and $t$.
\end{conjecture}
To support this Conjecture we perform a uniform random search over all pure input states in the finite dimensional Hilbert space. The restriction to search only among pure states is motivated by the following Lemmas. 

\begin{lemma} 
Given a self adjoint operator $H: \mathbb{C}^{K+1} \to \mathbb{C}^{K+1}$, 
we can always decompose a density operator $\rho$ on $\mathbb{C}^{K+1}$ satisfying a linear constraint ${\rm Tr}(\rho H)=N$, in terms of pure states  $|\psi_k\rangle$
satisfying the same constraint, i.e. ${\rm Tr}(|\psi_k\rangle\langle\psi_k| H)=N$.
\end{lemma}

\proof{
Consider the spectral decomposition of $H=\sum \limits _j  h _j |j\rangle\langle j|$.
An arbitrary density 
operator represented in the $H$ eigenvectors basis
\begin{equation}\label{rho1}
\rho =\sum \limits _{i,j} r _{i,j} |i\rangle\langle j|, \quad r_{j,j}>0,\quad\sum \limits _j r _{j,j}=1,
\end{equation}
satisfies the constrain if ${\rm Tr}(\rho H)=\sum \limits _j h_j r_{{j,j}}=N$.
Decomposing $\rho$ in terms of pure states we have
\begin{equation}\label{rho2}
\rho =\sum \limits _k p _k |\psi_k\rangle\langle \psi_k|,\quad p _k>0,\quad  \sum \limits _k p _k=1,
\end{equation}
Comparing Eqs.\eqref{rho1} and \eqref{rho2}, we find that 
$\sum \limits _k p _k|\langle \psi_k|j\rangle |^2=r _{{j,j}}$.
If we take 
\begin{equation}\label{condi}
|\langle \psi_k|j\rangle|^2=r_{{j,j}}, \quad \forall k,
\end{equation}
it will result 
\begin{equation*}
{\rm Tr} ( |\psi_k\rangle\langle \psi_k| H)=\sum \limits _j h_j r_{{j,j}} =N, \quad \forall k.
\end{equation*}
Hence it is enough to determine the $|\psi_k\rangle$s from the condition \eqref{condi} to get a decomposition 
of $\rho$ in terms of pure states satisfying the same constraint. This is always possible, actually in infinite many ways.
Additionally we have the freedom in choosing the $p_k$s.}

\begin{lemma} 
The minimum output entropy of a quantum channel $\Phi$ acting on states $\rho$ on 
$\mathbb{C}^{K+1}$ satisfying the energy constraint \eqref{inenergy} is achieved on pure states.
\end{lemma}

\proof{
Assume that the minimum output entropy is achieved by the input state $\rho$ satisfying the energy constraint. Decomposing it in terms of pure states that satisfy the same energy constraint 
$\rho=\sum_k p_k |\psi_k\rangle\langle \psi_k|$, and using the concavity of von Neumann entropy \cite{MOhya}, we have
\begin{eqnarray*}
S(\Phi(\rho))&=&S\big(\sum_k p_k \Phi\left(|\psi_k\rangle\langle \psi_k|\right)\big)\\
&\geq&\sum_k p_k S\left(\Phi\left(|\psi_k\rangle\langle \psi_k|\right)\right).
\end{eqnarray*}
In the decomposition, let us denote the pure state with minimum output entropy by $|\psi_*\rangle$. Therefore we have:
\begin{equation*}
S\left(\Phi(\rho)\right)\geq S\left(\Phi\left(|\psi_*\rangle\langle \psi_*|\right)\right),
\end{equation*}
that is, the optimal input state must be pure.
}

To generate random pure input states in $K+1$-dimensional
Hilbert space, we employ the following parametrization
\begin{equation*}
\begin{aligned}
|\psi\rangle&=\sum_{n=0}^{K} \nu_n |n\rangle,\\
\nu_0&=\cos\theta_K,\quad
\nu_{n>0}=e^{i\phi_n}\cos\theta_{K-n}\prod_{l=K-n+1}^{K}\sin\theta_l.
\end{aligned}
\end{equation*}
Then, according to \cite{Karol}, it is enough to generate $\phi_{n\geq 1}\in[0,2\pi)$ from 
a uniform distribution $p(\phi_{n\geq 1})=\frac{1}{2\pi}$ and random
independent variables $\xi_n$ distributed uniformly in $[0,1]$
for $n=1,\ldots, K$ defining
\begin{equation*}
\theta_n:=\arcsin(\xi_n^{\frac{1}{2n}}).
\end{equation*}
However, due to the energy constraint \eqref{inenergy},
we should consider states satisfying
 $\sum_{n=0}^{K} n |\nu_n|^2=N$.
This imposes a functional relation among $\theta_n$s and so
among $\xi_n$s, which can be written as:
 $\xi_{K}=g(\xi_1,\xi_2,\ldots\xi_{K-1};N)$.
Therefore we should generated $K-1$ random variables with the
following modified probability distribution function
\begin{equation*}
\begin{aligned}
\tilde{p}(\xi_1,\ldots, \xi_{K-1})
={\cal C}\int d\xi_{K} p(\xi_1,\ldots,\xi_{K})
\delta\left(\xi_{K}-g\right),
\end{aligned}
\end{equation*}
being ${\cal C}$ a normalization factor and
$p(\xi_1,\ldots,\xi_{K})$ the probability distribution
function for the variables $\xi_1,\ldots,\xi_{K}$.
Since these are chosen independently and with a standard uniform
distribution in $[0,1]$, we conclude that we should generate
$\xi_1,\ldots,\xi_{K-1}$ according to
$ \tilde{p}(\xi_1,\ldots, \xi_{K-1})=p(\xi_1,\ldots, \xi_{K-1})=1$,
and pick $\xi_{K}$ as
\begin{equation*}
\begin{aligned}
&\xi_{K}=g(\xi_1,\ldots, \xi_{K-1};N)\\
&=\frac{N}{\left[1+\xi_{K-1}^{1/{(K-1)}}\left(1+\xi_{K-2}^{1/(K-2)}\left(1+\cdots\xi_2^{1/2}\left(1+\xi_1\right)\right)\right)\right]}.\nonumber
\end{aligned}
\end{equation*}
In our $4$-dimensional example with $N=0.6$ the search over
$10^5$ states, generated as explained above, confirms the
statement of Conjecture \ref{conj:bin+magic}.

\vspace{-0.8cm}

\section{Conclusion}\label{sec:conclu}

We have opened an avenue for studying,
from an information theoretic point of view,
continuous quantum channels beyond the usual restriction of
Gaussianity.
Actually we have proposed a model of non-Gaussian quantum
channel that stems from a master equation accounting for two
processes, amplitude damping and dephasing. 
Its physical relevance relies on the fact that
amplitude damping and dephasing are
applied in many concrete discussions to model noise of quantum
information processing with single mode light field, vibration
phonon mode, or excitonic wave, see e.g. \cite{Lidar}.

Then, the first question that arises is how much the introduced
channel deviates from Gaussianity. Arguably this depends on the
parameter $\epsilon$, however an exact quantification would be
in order, maybe in a fashion similar to what has been done for
non-Gaussian states \cite{Konrad}. This could also shed light on
the choice of optimal input states for communication tasks.
Here we found input states approaching zero output
entropy, while respecting the input energy constraint.
They consist in the superposition of two number states the
farthest away one from the other. In truncated Hilbert space,
the minimum output entropy remains finite and optimal input
states are conjectured to be binomial states beside
superposition of two number states, depending on the values of
the channel's parameters. This is corroborated by numerical
results. The study performed in truncated Hilbert space is
justified by the fact that in realistic physical situations is
hard to fully exploit the infinite dimensionality of the space
$L^2(\mathbb{R})$.

As further development one could address the issue of additivity
of output entropy for two copies of the channel and then
eventually of multiple copies. This would be motivated by the
additivity of the classical capacity
deriving from the additivity of the minimum output entropy
\cite{Shor}.

Although challenging, 
the introduced map leaves concrete hopes for characterizing its
(product states) classical capacity which implies finding the
optimal input ensemble of states maximizing the Holevo chi
quantity \cite{Holevo}.

\vspace{-0.8cm}

\acknowledgments
S. M. would like to thank the Sharif University of Technology for
kind hospitality during the final stage of this work.



\end{document}